\documentclass[cameraready]{Interspeech}

\title{Gumbel-BEARD: Automatic Layer Selection for Self-Supervised Adaptation of Whisper in Low-Resource Domains}

\author[affiliation={1},orcid=0009-0006-3119-6159]{Zilai}{Wang}
\author[affiliation={1},orcid=0009-0005-9726-6597]{Natarajan Balaji }{Shankar}
\author[affiliation={1},orcid=0009-0009-5841-6846]{Mohan}{Shi}
\author[affiliation={1},orcid=0009-0005-2488-3770]{Kaiyuan}{Zhang}
\author[affiliation={1},orcid=0009-0004-1406-503X]{Abeer}{Alwan}

\address{
    $^1$ University of California, Los Angeles, USA
}

\email{ \{zilaiwang2001,balaji1312,shimohan,kaiyuanzhang\}@ucla.edu, alwan@ee.ucla.edu}


\usepackage{comment}

\usepackage{multirow}

\begin{document}

\maketitle


\begin{abstract}
Speech foundation models often struggle in low-resource domains due to domain mismatch and data scarcity. We propose Gumbel-BEARD, a domain adaptation framework that automates Whisper encoder layer selection via an end-to-end trainable hard Gumbel-Softmax selector. It enables self-supervised adaptation with a BEST-RQ objective that dynamically adapts to target acoustic characteristics without manual tuning. Experiments on the MyST child speech corpus demonstrate efficiency and scalability: with 10\,h of labeled data for fine-tuning, our method matches a fully supervised baseline trained on the complete 133\,h labeled set. We establish new state-of-the-art word error rates (WERs) of 8.21\% using Whisper-medium on MyST and 11.06\% using Whisper-small on the OGI Spontaneous dataset. Evaluation on CORAAL further confirms robustness to adult dialectal domain shifts, with up to 6\% relative WER reduction, highlighting the generalizability of our approach to diverse low-resource conditions.
\end{abstract}

\section{Introduction}
\label{sec:introduction}

Recent advancements in automatic speech recognition (ASR) have been driven by deep neural networks trained on large-scale datasets, yielding strong end-to-end models such as OpenAI Whisper~\cite{radford2023robust}, Meta SeamlessM4T~\cite{seamless}, NVIDIA Canary~\cite{canary}, and OWSM~\cite{owsm}. However, these models suffer significant performance degradation in low-resource domains, where domain mismatch and data scarcity remain critical challenges~\cite{mlsuperb}.

To mitigate this, researchers have widely adopted data augmentation strategies to artificially expand training distributions~\cite{jaitly2013vocal,park2019specaugment,KoPPK15,BaasK22}, alongside Parameter-Efficient Fine-Tuning (PEFT) approaches~\cite{HoulsbyGJMLGAG19,LiL20,HuSWALWWC22,LiuJFTDY022} for efficient model adaptation. Knowledge transfer from high-resource domains has also been extensively explored~\cite{ShivakumarG20,sinha2025beyond,RollandACS22,shankar2025selective,NagasawaOI25,shankara2026compositional}, while feature fusion methods~\cite{BerrebbiSYLA022,SrivastavaSC024,ChiuWH0W24,wang2026mind} aim to exploit complementary information across diverse speech representations. Despite their effectiveness, these strategies often rely on labeled data, which remains scarce in many target domains.

Consequently, Unsupervised Domain Adaptation (UDA) has gained traction as a means of leveraging unlabeled audio to adapt model representations~\cite{Kahn0H20,HwangSHS22,HwangMHSGQSSBH22,HuCYQCCZ24,ShankarFA24,shi2025comparing}. The recent BEARD framework~\cite{bagat2025best} proposed the integration of self-supervised learning (SSL) objectives for adapting Whisper to new domains. However, applying such objectives to deep Transformer architectures is non-trivial, as determining the optimal intermediate encoder layer at which to apply the masked self-supervised prediction loss (the \emph{prediction layer}) remains largely heuristic and computationally prohibitive, often requiring an expensive search over candidate layers. Other techniques such as weighted sums across layers are differentiable but add computation and can blur contributions by mixing abstraction levels~\cite{ShihH24}. This limitation is particularly acute in child ASR, where recognition performance lags significantly behind adult benchmarks~\cite{fan2024benchmarking}. Shorter vocal tracts, higher fundamental frequency, acoustic variability, and disfluencies~\cite{lee1999acoustics} create a complex distribution shift that heuristic fixed-layer adaptation methods are ill-suited to handle, motivating the need for automated solutions capable of efficiently locating informative encoder representations.

To address these challenges, we propose Gumbel-BEARD, an automated domain adaptation framework based on the Gumbel-Softmax estimator~\cite{jang2017categorical}, which dynamically selects a prediction layer at each optimization step during self-supervised adaptation. By making layer selection an end-to-end trainable process, our framework explores diverse encoder representations using exclusively unlabeled data, eliminating the need for manual layer search.
The main contributions of this work are:
\begin{itemize}
    \item We propose Gumbel-BEARD, an automatic layer selection framework integrating hard Gumbel-Softmax with a self-supervised objective. 

    \item We establish state-of-the-art (SOTA) WERs on two child speech corpora: 8.21\% on MyST~\cite{ward2011my} using Whisper-medium and 11.06\% on the OGI Spontaneous test set~\cite{ogi_kids} using Whisper-small. 
    \item We demonstrate that Gumbel-BEARD generalizes to adult dialectal speech (CORAAL~\cite{kendall2023coraal}), with up to 6\% relative WER reduction, indicating its effectiveness across diverse acoustic and linguistic domain shifts.\footnote{\url{https://github.com/Zilai-WANG/Gumbel_Beard}}
\end{itemize}

\begin{figure*}[t]
    \centering
    \includegraphics[width=0.8\textwidth]{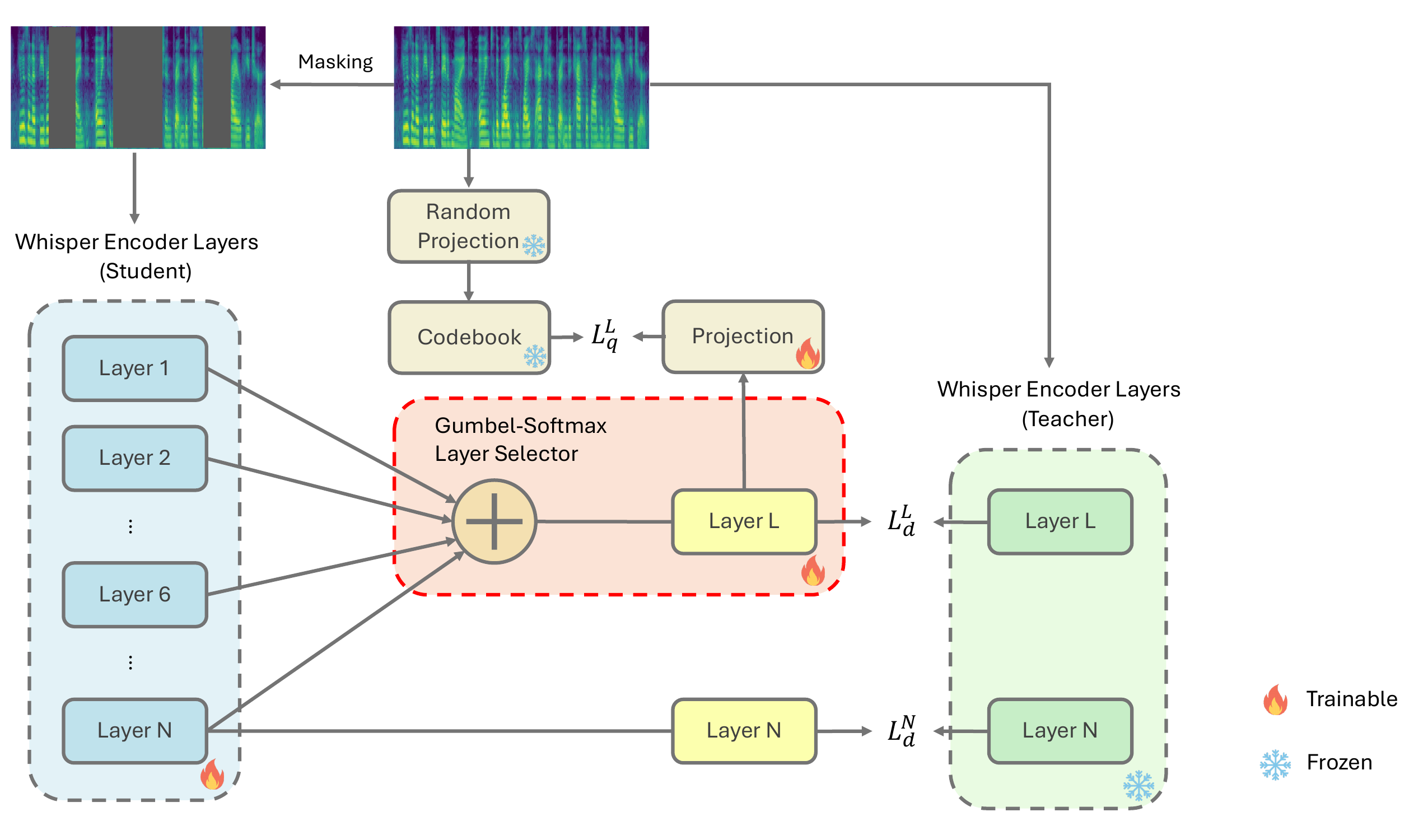} 
    \caption{Overview of the proposed Gumbel-BEARD framework. A (hard) Gumbel-Softmax layer selector selects a prediction layer $L$ from the student encoder at each optimization step. The model is optimized by aligning the selected representation with BEST-RQ discrete codebook targets via the quantization loss $\mathcal{L}_q^L$, and with frozen teacher representations via inner and output distillation losses ($\mathcal{L}_d^L$ and $\mathcal{L}_d^N$).}
    \label{fig:gumbel-beard}
\end{figure*}

\section{Methods}
\label{sec:methods}

\subsection{Background: The BEARD Framework}
\label{ssec:beard}

BEARD~\cite{bagat2025best} adapts Whisper through a two-stage procedure. In the first stage, the Whisper encoder is adapted on unlabeled data using a combination of self-supervised learning and distillation, while the decoder is excluded from training. For the self-supervised objective, BEARD adopts BEST-RQ~\cite{chiu2022self}, where a frozen random projection maps speech features to a codebook, yielding discrete targets for a masked prediction task at a designated prediction layer $L$. To preserve pre-trained knowledge, a dual distillation constraint maximizes cosine similarity between the student and a frozen teacher at both the prediction layer $L$ (inner distillation) and the final encoder layer $N$ (output distillation). The stage-one objective is:
\begin{equation}
    \mathcal{L} = \mathcal{L}_q^L + \lambda \mathcal{L}_d^L + \beta \lambda \mathcal{L}_d^N
\end{equation}
where $\mathcal{L}_q^L$ is the BEST-RQ quantization loss, and $\mathcal{L}_d^L$, $\mathcal{L}_d^N$ denote the inner and output distillation losses, respectively. In the second stage, the adapted encoder is recoupled with the decoder and jointly fine-tuned on limited labeled data. In the standard BEARD framework, the prediction layer $L$ is a fixed hyperparameter, requiring manual tuning prior to adaptation.


\subsection{Proposed Method: Gumbel-BEARD}
\label{ssec:gumbel_beard}

While BEARD demonstrates the utility of self-supervised domain adaptation, it has two key limitations. 
First, a single fixed prediction layer may be suboptimal across diverse low-resource domains, since the appropriate level of feature abstraction varies with acoustic conditions~\cite{ChenH0A0RLTA25}. 
Second, determining this layer typically relies on brute-force search, which becomes computationally prohibitive as model size increases.

To address these limitations, we propose Gumbel-BEARD, which replaces the fixed prediction layer with a hard Gumbel-Softmax selector that chooses the prediction layer at each optimization step. 
This enables the model to leverage different levels of encoder abstraction throughout training, as illustrated in Figure~\ref{fig:gumbel-beard}.

\subsubsection{Hard Gumbel-Softmax Selection}
We introduce a learnable parameter $\boldsymbol{\alpha} \in \mathbb{R}^N$ representing the unnormalized log-probabilities of selecting each encoder layer. To enable discrete selection while preserving gradient-based optimization, we employ the Gumbel-Softmax estimator~\cite{jang2017categorical}. Soft probabilities $y_i$ are computed via the Gumbel-Softmax relaxation:
\begin{equation}
    y_i = \frac{\exp((\alpha_i + g_i) / \tau)}{\sum_{j=1}^N \exp((\alpha_j + g_j) / \tau)}
\end{equation}
where $g_i \sim \text{Gumbel}(0,1)$ are i.i.d.\ samples and $\tau$ is the temperature. Instead of a \textit{soft} variant, which computes a weighted sum of all layers via these probabilities, we use a \textit{hard} variant that applies an argmax to yield a discrete one-hot selection:
\begin{equation}
    \mathbf{z}_{\text{hard}} = \mathrm{OneHot}(\operatorname*{argmax}_i(y_i))
\end{equation}
Since argmax is non-differentiable, gradients flow through the soft probabilities $\mathbf{y}$ via the Straight-Through Estimator (STE)~\cite{BengioLC13}, enabling end-to-end optimization of $\boldsymbol{\alpha}$. The student prediction representation is then extracted via the hard selection vector, $\hat{\mathbf{H}}_S = \sum_{i=1}^N (\mathbf{z}_{\text{hard}})_i \mathbf{H}_S^i$. Following BEARD, the inner distillation loss $\mathcal{L}_d^L$ is the cosine similarity between this dynamically selected student representation and the corresponding hidden state of the frozen teacher encoder at the same layer index $L$.

\subsubsection{Temperature Annealing}
The temperature $\tau$ governs the exploration-exploitation trade-off. Initialized to a high value, $\tau$ approximates a uniform distribution, ensuring all encoder layers are adequately explored in early training. As training progresses, $\tau$ is annealed to sharpen the distribution, increasingly concentrating selection on fewer layers that are most effective for the BEST-RQ objective.

\section{Experiments}
\label{sec:experiments}

\subsection{Datasets}
\label{ssec:datasets}

To evaluate the efficacy of our proposed method on domain shifts, we conduct experiments on three distinct corpora representing child speech and dialectal variations.

\begin{itemize}
    \item \textbf{MyST~\cite{ward2011my}:} A large corpus of conversational child speech from students (grades 3--5) interacting with a virtual science tutor. Of the total 448\,h, only 240\,h are transcribed; filtering following~\cite{attia2023kid} retains 133/21/25\,h for train/dev/test, with 1\,h and 10\,h subsets randomly sampled from the training partition for limited-supervision experiments. The remaining 208\,h of untranscribed audio is used for self-supervised adaptation.

    \item \textbf{OGI Kids~\cite{ogi_kids}:} Contains scripted and spontaneous child speech. Following~\cite{cmu_benchmark}, spontaneous data is split into 22/2/7\,h for train/dev/test, with the test set stratified into three age groups (4--7, 8--10, 11--15). The scripted 48\,h training subset~\cite{fan2024benchmarking} is combined with spontaneous 22\,h (70\,h total) for adaptation, while fine-tuning uses the spontaneous partition only.

    \item \textbf{CORAAL~\cite{kendall2023coraal}:} Sociolinguistic interviews in African American Language. We use six subsets (ATL, LES, DCA, DCB, DTA, PRV; 137\,h) for training, and hold out ROC (13\,h) and VLD (12\,h) for development and testing, ensuring speaker and regional disjointness. Utterances are trimmed to retain only interviewee speech and capped at 30\,s. The 137\,h training partition is used as unlabeled data for self-supervised adaptation and as labeled data for fine-tuning.
\end{itemize}


\subsection{Models}
\label{ssec:models}

 We adopt Whisper as the backbone to enable a controlled comparison with our primary baseline, BEARD~\cite{bagat2025best}. Specifically, we investigate:
\begin{itemize}
    \item \textbf{Whisper-small} (244M parameters): Comprises 12 encoder and 12 decoder layers with hidden dimension 768.
    \item \textbf{Whisper-medium} (769M parameters): Comprises 24 encoder and 24 decoder layers with hidden dimension 1024.
\end{itemize}

\subsection{Experimental Setup}
\label{ssec:setup}

\subsubsection{Baselines}
We compare Gumbel-BEARD against three competitive baselines to validate data efficiency and adaptation performance:

\begin{enumerate}
    \item \textbf{Supervised Fine-Tuning (SFT):} Following~\cite{fan2024benchmarking}, the pre-trained Whisper model is directly fine-tuned on the labeled target data without any prior self-supervised adaptation.
    
    \item \textbf{Standard BEARD~\cite{bagat2025best}:} The original framework with a fixed prediction layer determined via manual search. All other self-supervised adaptation hyperparameters follow the default configuration of~\cite{bagat2025best}, and fine-tuning hyperparameters follow~\cite{fan2024benchmarking}.
    
    \item \textbf{Pseudo-Labeling (PL):} A semi-supervised baseline. We employ Whisper-large-v3 to generate pseudo-transcriptions for unlabeled data. To ensure quality, we apply text normalization to remove formatting artifacts. The model is then fine-tuned on the combined dataset (Ground Truth + Pseudo-Labeled).
\end{enumerate}

\subsubsection{Gumbel-BEARD Implementation}
\noindent\textbf{Self-supervised adaptation:} Layer selection is parameterized by a learnable logit vector $\boldsymbol{\alpha} \in \mathbb{R}^N$, initialized to zero for a uniform prior over all encoder layers. The temperature $\tau$ is initialized at $5.0$ and annealed linearly to $0.1$ throughout training. All other hyperparameters follow the BEARD baseline: learning rate $1\times10^{-4}$, batch size 32, 1 epoch, $\lambda=0.5$, $\beta=0.1$, and codebook size 2048. We apply this single configuration unchanged across all datasets and model sizes.

\noindent\textbf{Fine-tuning:} The adapted encoder is reintegrated with the original Whisper decoder and jointly fine-tuned on available labeled data following the BEARD protocol, with hyperparameters from~\cite{fan2024benchmarking}. All experiments are conducted on a single NVIDIA RTX 5090 GPU. Statistical significance is assessed using the Matched-Pairs Sentence-Segment Word Error (MAPSSWE) test ($p < 0.05$) implemented in the NIST SCTK toolkit~\cite{NIST-SCTK}.

\noindent\textbf{Ablation: soft vs.\ hard layer selection.}
We compare a \textit{soft} variant, which computes a weighted sum of all layers via the Gumbel-Softmax probabilities, against a \textit{hard} variant that applies an argmax to yield a discrete one-hot selection. Results are shown in Section~\ref{sec:results}.

\subsection{Canonical Correlation Analysis}
Following~\cite{pasad2023comparative}, we quantify representational similarity using Canonical Correlation Analysis (CCA)~\cite{hotelling1992relations}, which is invariant to invertible linear transformations. CCA identifies linear projections that maximize the correlation between two representations $X$ and $Y$:
\begin{equation}
    v_1, w_1 = \arg\max_{v,w} \mathrm{corr}(v^\top X, w^\top Y).
\end{equation}
Specifically, we employ Projection-Weighted CCA (PWCCA)~\cite{morcos2018insights}, which aggregates canonical correlations into a robust scalar similarity score.

\section{Results}
\label{sec:results}

\subsection{Comparison with Baselines on MyST}
\label{ssec:comparison_myst}
Table~\ref{tab:comparison_myst} reports WER on the MyST test set with the Whisper-small backbone. Gumbel-BEARD (hard selection) consistently outperforms all baselines across labeled data budgets, with statistically significant gains ($p < 0.05$) over SFT. With only 10\,h of labeled data, it attains 9.35\% WER, nearly matching the SFT baseline trained on the full 133\,h set (9.34\%); on the full dataset it reaches 8.51\%, surpassing standard BEARD (8.73\%). The soft variant underperforms hard selection at every budget, indicating the benefit of discrete layer routing. We attribute this to gradient interference: soft selection mixes representations from different abstraction levels, whereas hard selection commits to a single representation per step, yielding a cleaner training signal. Hereafter, Gumbel-BEARD denotes the hard variant. In addition to improving recognition performance, Gumbel-BEARD reduces adaptation cost to approximately 1 GPU-hour on Whisper-small, compared to roughly 12 GPU-hours for the exhaustive layer search required by BEARD. Pseudo-labeling is even more computationally expensive due to the additional inference required for pseudo-transcription generation.



\begin{table}[ht]
    \centering
    \caption{WER (\%) on the MyST test set using Whisper-small across varying amounts of labeled fine-tuning data. Zero-shot performance is provided as reference. \textbf{Bold} indicates best results and $^*$ denotes statistically significant improvement ($p < 0.05$) over SFT.}
    \vspace{-8pt}
    \label{tab:comparison_myst}
    \setlength{\tabcolsep}{8pt}
    \begin{tabular}{l c c c}
        \toprule
        \multirow{2}{*}{\textbf{Method}} & \multicolumn{3}{c}{\textbf{Labeled Data Size}} \\
        \cmidrule(lr){2-4}
         & \textbf{1\,h} & \textbf{10\,h} & \textbf{Full} \\
        \midrule
        Zero-shot & \multicolumn{3}{c}{13.40 (no fine-tuning)} \\
        \midrule
        Baseline (PL)             & 12.45 & 11.66 & 9.63 \\
        Baseline (SFT)            & 10.64 & 9.94  & 9.34 \\
        Baseline (BEARD)~\cite{bagat2025best}          & 10.31 & 9.44  & 8.73 \\
        \midrule
        Proposed (Gumbel-BEARD)     &       &       &      \\
        \quad w/ Soft selection     & 10.65 & 9.62  & 8.76 \\
        \quad w/ Hard selection     & \textbf{10.18}\rlap{*} & \textbf{9.35}\rlap{*} & \textbf{8.51}\rlap{*} \\
        \bottomrule
    \end{tabular}
\end{table}


\subsection{Scalability to a Larger Architecture}
\label{ssec:scalability}

To verify that Gumbel-BEARD extends beyond smaller models, we scale experiments to the Whisper-medium architecture, which features a deeper 24-layer encoder. As shown in Table~\ref{tab:scalability}, Gumbel-BEARD consistently outperforms the SFT baseline across all labeled data budgets. The PL baseline is omitted due to its inferior performance relative to other baselines, and the standard BEARD baseline is excluded as exhaustive layer search becomes computationally prohibitive at this model scale.

Fine-tuning on the full dataset yields 8.21\% WER, improving over the previously best reported MyST test WER of 8.50\%~\cite{fan2024benchmarking} achieved by the substantially larger Parakeet model (1.1B parameters)~\cite{RekeshKKMNHHPKBG23}. These results confirm that the automated layer selection mechanism scales effectively to deeper and more complex Transformer architectures.

\begin{table}[ht]
    \centering
    \caption{WER (\%) on the MyST test set using Whisper-medium, comparing the SFT baseline and Gumbel-BEARD across varying amounts of labeled fine-tuning data.}
    \vspace{-8pt}
    \label{tab:scalability}
    \setlength{\tabcolsep}{8pt}
    \begin{tabular}{l c c c}
        \toprule
        \multirow{2}{*}{\textbf{Method}} & \multicolumn{3}{c}{\textbf{Labeled Data Size}} \\
        \cmidrule(lr){2-4}
         & \textbf{1\,h} & \textbf{10\,h} & \textbf{Full} \\
        \midrule
         Zero-shot & \multicolumn{3}{c}{13.10 (no fine-tuning)} \\
        \midrule
        Baseline (SFT)          & 9.56 & 9.19 & 8.86 \\
        Proposed (Gumbel-BEARD) & \textbf{9.15}\rlap{*} & \textbf{8.88}\rlap{*} & \textbf{8.21}\rlap{*} \\
        \bottomrule
    \end{tabular}
\end{table}
\vspace{-10pt}
\subsection{Cross-Domain Transferability}
\label{ssec:cross_domain}

We evaluate cross-domain transferability of Gumbel-BEARD by performing self-supervised adaptation on the unlabeled MyST corpus, followed by supervised fine-tuning and evaluation on OGI Spontaneous. We utilize the Whisper-small backbone, as larger model sizes have been shown to overfit on low-resource child ASR datasets such as OGI Spontaneous~\cite{cmu_benchmark}, which we observed in preliminary experiments. Results are presented in Table~\ref{tab:transferability}.
In-domain OGI adaptation achieves a WER of 11.06\% on the OGI Spontaneous test set, surpassing the previous best reported WER~\cite{cmu_benchmark}. Adapting on out-of-domain MyST yields a comparable overall WER of 11.15\%, outperforming the SFT baseline and achieving near parity with in-domain OGI adaptation. Within the 8--10 age group, the cross-domain model achieves the lowest WER of 10.19\%, which may reflect the demographic overlap between MyST (grades 3--5) and this age cohort, as well as the larger unlabeled adaptation set (208\,h). Results support effective cross-domain transfer to target populations with similar acoustics.
\begin{table}[ht]
    \centering
    \caption{Cross-domain transferability on the OGI Spontaneous test set (Whisper-small). Out-of-domain MyST adaptation is compared against SFT baseline and in-domain OGI adaptation.}
    \vspace{-8pt}
    \label{tab:transferability}
    \setlength{\tabcolsep}{4pt}
    \begin{tabular}{l @{\hskip 4pt} c c c c}
        \toprule
        \textbf{Self-supervised} & \textbf{WER} & \multicolumn{3}{c}{\textbf{WER by Age Group}} \\
        \cmidrule(lr){3-5}
        \textbf{Adaptation Data}  & \textbf{(Overall)} & \textbf{4--7} & \textbf{8--10} & \textbf{11--15} \\
        \midrule
        None (Zero-shot)    & 26.29 & 36.63 & 28.18 & 20.49 \\
        None (SFT Baseline) & 11.57 & 17.60 & 10.90 & 9.26 \\
        OGI (In-Domain)     & \textbf{11.06}\rlap{*} & \textbf{17.43} & 10.42\rlap{*} & \textbf{8.59}\rlap{*} \\
        MyST (Cross-Domain) & 11.15\rlap{*} & 17.78 & \textbf{10.19}\rlap{*} & 8.74\rlap{*} \\
        \bottomrule
    \end{tabular}
\end{table}


\subsection{Evaluation on Dialectal Speech}
\label{ssec:coraal}
To assess the robustness of Gumbel-BEARD beyond child speech, we evaluate its performance on the CORAAL dataset, which presents a distinct domain shift in the form of dialectal and sociolinguistic variation. Table~\ref{tab:coraal} presents WER results across Whisper-small and Whisper-medium architectures, compared against the SFT baseline.

Gumbel-BEARD improves over the SFT baseline on both development and test splits across both architectures. On Whisper-small, the proposed method reduces test WER from 11.70\% to 11.01\%. Performance further improves with model capacity: Whisper-medium achieves a test WER of 9.25\% compared to 9.81\% for SFT. These results demonstrate that the automated layer selection mechanism generalizes to sociolinguistic domain shifts that are acoustically and linguistically distinct from child speech.

\begin{table}[ht]
    \centering
    \caption{WER (\%) on the CORAAL dataset for Whisper-small and Whisper-medium, comparing zero-shot, SFT baseline, and Gumbel-BEARD on development and test splits.}
    \vspace{-8pt}
    \label{tab:coraal}
    \setlength{\tabcolsep}{4pt}
    \begin{tabular}{l c c c c}
        \toprule
        & \multicolumn{2}{c}{\textbf{Whisper-small}} & \multicolumn{2}{c}{\textbf{Whisper-medium}} \\
        \cmidrule(lr){2-3} \cmidrule(lr){4-5}
        \textbf{Method} & \textbf{Dev} & \textbf{Test} & \textbf{Dev} & \textbf{Test} \\
        \midrule
        Zero-shot      & 13.43 & 18.69 & 13.79 & 17.09 \\
        Baseline (SFT) & 7.30  & 11.70 & 6.51  & 9.81 \\
        Gumbel-BEARD  & \textbf{7.19} & \textbf{11.01}\rlap{*} & \textbf{6.10}\rlap{*} & \textbf{9.25}\rlap{*} \\
        \bottomrule
    \end{tabular}
\end{table}


\subsection{Layer Representation Analysis}
\label{ssec:pwcca}

\begin{figure}[ht]
    \centering
    \includegraphics[width=.9\columnwidth]{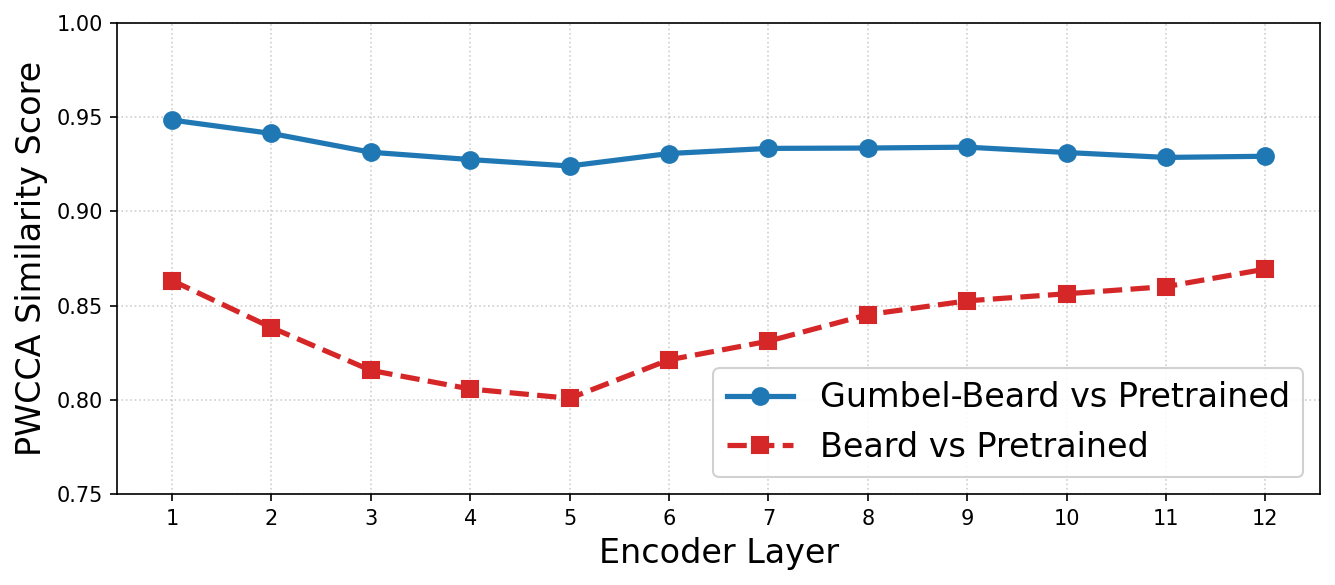} 
    \vspace{-0.2cm}
    \caption{Layer-wise PWCCA similarity between the adapted and original Whisper-small encoder. Gumbel-BEARD (blue) preserves more original representations than BEARD (red).}
    \label{fig:pwcca}
    \vspace{-15pt}
\end{figure}

We analyze the representational similarity between the adapted and original Whisper-small encoders using PWCCA on the MyST dev set. As illustrated in Figure~\ref{fig:pwcca}, Gumbel-BEARD maintains higher similarity scores across all encoder layers compared to the standard BEARD baseline, with a smoother layer-wise trajectory. This suggests that the dynamic selection strategy adapts to the target domain while better preserving the original knowledge. We hypothesize that this stems from Gumbel-BEARD applying the inner distillation loss dynamically across layers over training: broad exploration in early, high-temperature steps regularizes all encoder layers, before the selector concentrates on the medium layers as the temperature is annealed. In contrast, BEARD restricts the regularization signal to a single fixed prediction layer throughout adaptation.

\section{Conclusion}
\label{sec:conclusion}
We introduce Gumbel-BEARD, a domain adaptation framework that automates Whisper encoder layer selection via an end-to-end trainable hard Gumbel-Softmax layer selector. Our experiments establish state-of-the-art WERs of 8.21\% on MyST and 11.06\% on the OGI Spontaneous test set, with generalization to adult dialectal speech (CORAAL). PWCCA analysis suggests that dynamic layer selection better preserves the original knowledge throughout adaptation. Future work will explore extending this approach to other encoder-decoder Transformer architectures, including speech LLMs. We recently validated our method on Canary-Qwen \cite{nvidia_canary_qwen_25b} and observed similar improvements.

\section{Acknowledgements}
\label{sec:ack}
This research is supported in part by the National Science Foundation (NSF) and the Institute of Education Sciences (IES), U.S. Department of Education (DoE), through Grant R305C240046 to the U. at Buffalo. The opinions expressed are those of the authors and do not represent views of the IES, DoE, or the NSF. 

\section{Generative AI Use Disclosure}
\label{sec:genai_disclosure}

During the preparation of this manuscript, the authors utilized ChatGPT (GPT-5.2) exclusively for language editing, including proofreading and enhancing readability. All technical content, experimental design, results, and conclusions were independently developed and verified by the authors. Following the use of this tool, the authors carefully reviewed and revised the manuscript, assuming full responsibility for the final content. Generative AI was not used to generate any substantive scientific material and is not credited as an author.

\bibliographystyle{IEEEtran}
\bibliography{mybib}

@inproceedings{radford2023robust,
  title={Robust speech recognition via large-scale weak supervision},
  author={A. Radford and others},
  booktitle={Proc. ICML},
  year={2023},
}

@article{seamless,
  author       = {Lo{\"{\i}}c Barrault and others},
  title        = {Seamless: Multilingual Expressive and Streaming Speech Translation},
  journal      = {CoRR},
  volume       = {abs/2312.05187},
  year         = {2023},
  url          = {https://doi.org/10.48550/arXiv.2312.05187},
  doi          = {10.48550/ARXIV.2312.05187},
  eprint       = {2312.05187},
  eprinttype   = {arXiv}
}

@inproceedings{canary,
  author       = {Krishna C. Puvvada and others},
  title        = {Less is More: Accurate Speech Recognition {\&} Translation without
                  Web-Scale Data},
  booktitle    = {{INTERSPEECH}},
  year         = {2024}
}

@inproceedings{owsm,
  author       = {Yifan Peng and others},
  title        = {{OWSM} v3.1: Better and Faster Open Whisper-Style Speech Models based
                  on E-Branchformer},
  booktitle    = {{INTERSPEECH}},
  year         = {2024}
}

@inproceedings{mlsuperb,
  author       = {Jiatong Shi and others},
  title        = {{ML-SUPERB} 2.0: Benchmarking Multilingual Speech Models Across Modeling
                  Constraints, Languages, and Datasets},
  booktitle    = {{INTERSPEECH}},
  year         = {2024}
}

@article{lee1999acoustics,
  title={Acoustics of children’s speech: Developmental changes of temporal and spectral parameters},
  author={S. Lee and others},
  journal={The Journal of the Acoustical Society of America},
  volume={105},
  number={3},
  pages={1455--1468},
  year={1999}
}

@article{jaitly2013vocal,
  title={Vocal tract length perturbation (VTLP) improves speech recognition},
  author={Jaitly, Navdeep and Hinton, Geoffrey E},
  journal={Proc. ICML Workshop on Deep Learning for Audio, Speech and Language},
  year={2013}
}

@inproceedings{park2019specaugment,
  author       = {Daniel S. Park and others},
  title        = {SpecAugment: {A} Simple Data Augmentation Method for Automatic Speech
                  Recognition},
  booktitle    = {{INTERSPEECH}},
  year         = {2019},
}

@inproceedings{KoPPK15,
  author       = {Tom Ko and others},
  title        = {Audio augmentation for speech recognition},
  booktitle    = {{INTERSPEECH}},
  year         = {2015}
}

@inproceedings{BaasK22,
  author       = {Matthew Baas and
                  Herman Kamper},
  title        = {Voice Conversion Can Improve {ASR} in Very Low-Resource Settings},
  booktitle    = {{INTERSPEECH}},
  year         = {2022}
}

@inproceedings{HoulsbyGJMLGAG19,
  author       = {Neil Houlsby and others},
  title        = {Parameter-Efficient Transfer Learning for {NLP}},
  booktitle    = {{ICML}},
  year         = {2019}
}

@inproceedings{HuSWALWWC22,
  author       = {Edward J. Hu and others},
  title        = {LoRA: Low-Rank Adaptation of Large Language Models},
  booktitle    = {{ICLR}},
  year         = {2022}
}

@inproceedings{LiuJFTDY022,
  author       = {Xiao Liu and others},
  title        = {P-Tuning: Prompt Tuning Can Be Comparable to Fine-tuning Across Scales
                  and Tasks},
  booktitle    = {{ACL} {(2)}},
  year         = {2022}
}

@inproceedings{LiL20,
  author       = {Xiang Lisa Li and
                  Percy Liang},
  title        = {Prefix-Tuning: Optimizing Continuous Prompts for Generation},
  booktitle    = {{ACL/IJCNLP} {(1)}},
  year         = {2021}
}

@article{ShivakumarG20,
  author       = {Prashanth Gurunath Shivakumar and
                  Panayiotis G. Georgiou},
  title        = {Transfer learning from adult to children for speech recognition: Evaluation,
                  analysis and recommendations},
  journal      = {Comput. Speech Lang.},
  volume       = {63},
  pages        = {101077},
  year         = {2020}
}

@inproceedings{RollandACS22,
  author    = {Thomas Rolland and others},
  title     = {Multilingual Transfer Learning for Children Automatic Speech Recognition},
  booktitle = {{LREC}},
  year      = {2022}
}

@inproceedings{shankar2025selective,
author={Shankar, Natarajan Balaji and others},
  title     = {Selective Attention Merging for Low Resource Tasks: A Case Study of Child {ASR}},
  booktitle = {{ICASSP}},
  year      = {2025}
}

@article{shankara2026compositional,
  title   = {Compositional domain adaptation for automatic speech recognition with headwise selective attention merging},
  journal = {Computer Speech \& Language},
  pages   = {102012},
  year    = {2026},
  issn    = {0885-2308},
  doi     = {https://doi.org/10.1016/j.csl.2026.102012},
  author  = {Natarajan Balaji Shankar and others}
}

@inproceedings{NagasawaOI25,
  author       = {Haruki Nagasawa and
                  Shinta Otake and
                  Shinji Iwata},
  title        = {Task Vector Arithmetic for Low-Resource {ASR}},
  booktitle    = {{ICASSP}},
  year         = {2025}
}

@inproceedings{sinha2025beyond,
  title={Beyond traditional speech modifications: Utilizing self supervised features for enhanced zero-shot children asr},
  author={Sinha, Abhijit and Kathania, Hemant Kumar and Kurimo, Mikko},
  booktitle={{INTERSPEECH}},
  year={2025}
}

@inproceedings{BerrebbiSYLA022,
  author       = {Dan Berrebbi and others},
  title        = {Combining Spectral and Self-Supervised Features for Low Resource Speech
                  Recognition and Translation},
  booktitle    = {{INTERSPEECH}},
  year         = {2022}
}

@inproceedings{SrivastavaSC024,
  author       = {Tejes Srivastava and others},
  title        = {{EFFUSE:} Efficient Self-Supervised Feature Fusion for {E2E} {ASR}
                  in Low Resource and Multilingual Scenarios},
  booktitle    = {{INTERSPEECH}},
  year         = {2024}
}

@inproceedings{ChiuWH0W24,
  author       = {Sheng{-}Chieh Chiu and others},
  title        = {Learnable Layer Selection and Model Fusion for Speech Self-Supervised
                  Learning Models},
  booktitle    = {{INTERSPEECH}},
  year         = {2024}
}

@inproceedings{wang2026mind,
  title={Mind the Shift: Using Delta SSL Embeddings to Enhance Child ASR},
  author={Wang, Zilai and others},
  booktitle={{ICASSP}},
  year={2026}
}

@inproceedings{Kahn0H20,
  author       = {Jacob Kahn and
                  Ann Lee and
                  Awni Y. Hannun},
  title        = {Self-Training for End-to-End Speech Recognition},
  booktitle    = {{ICASSP}},
  year         = {2020}
}

@inproceedings{HwangSHS22,
  author       = {Dongseong Hwang and others},
  title        = {Pseudo Label Is Better Than Human Label},
  booktitle    = {{INTERSPEECH}},
  year         = {2022}
}

@inproceedings{HwangMHSGQSSBH22,
  author       = {Dongseong Hwang and others},
  title        = {Large-Scale {ASR} Domain Adaptation Using Self- and Semi-Supervised
                  Learning},
  booktitle    = {{ICASSP}},
  year         = {2022}
}

@inproceedings{HuCYQCCZ24,
  author       = {Yuchen Hu and others},
  title        = {Self-Taught Recognizer: Toward Unsupervised Adaptation for Speech
                  Foundation Models},
  booktitle    = {{NeurIPS}},
  year         = {2024}
}

@inproceedings{ShankarFA24,
  author       = {Natarajan Balaji Shankar and
                  Ruchao Fan and
                  Abeer Alwan},
  title        = {{SOA:} Reducing Domain Mismatch in {SSL} Pipeline by Speech Only Adaptation
                  for Low Resource {ASR}},
  booktitle    = {{ICASSP} Workshops},
  year         = {2024}
}

@inproceedings{
shi2025comparing,
title={Comparing Unsupervised and Supervised Semantic Speech Tokens: A Case Study of Child {ASR}},
author={Mohan Shi and others},
booktitle={IEEE ASRU Satellite Workshop-AI for Children's Speech and Language},
year={2025},
}

@inproceedings{bagat2025best,
  title = {BEST-RQ-Based Self-Supervised Learning for Whisper Domain Adaptation},
    author={Bagat, Rapha{\"e}l and Illina, Irina and Vincent, Emmanuel},
  booktitle = {{ICASSP}},
  year = {2026},
}

@inproceedings{ShihH24,
  author       = {Yi{-}Jen Shih and
                  David Harwath},
  title        = {Interface Design for Self-Supervised Speech Models},
  booktitle    = {{INTERSPEECH}},
  year         = {2024}
}

@inproceedings{chiu2022self,
  title={Self-supervised learning with random-projection quantizer for speech recognition},
  author={Chiu, Chung-Cheng and others},
  booktitle={{ICML}},
  year={2022},
}

@inproceedings{jang2017categorical,
  title={Categorical Reparameterization with Gumbel-Softmax},
  author={Jang, Eric and Gu, Shixiang and Poole, Ben},
  booktitle={{ICLR}},
  year={2017}
}

@article{BengioLC13,
  author       = {Yoshua Bengio and
                  Nicholas L{\'{e}}onard and
                  Aaron C. Courville},
  title        = {Estimating or Propagating Gradients Through Stochastic Neurons for
                  Conditional Computation},
  journal      = {CoRR},
  volume       = {abs/1308.3432},
  year         = {2013}
}

@inproceedings{pasad2023comparative,
  title     = {Comparative layer-wise analysis of self-supervised speech models},
  author    = {A. Pasad and others},
  booktitle = {{ICASSP}},
  year      = {2023},
}

@incollection{hotelling1992relations,
  title     = {Relations between two sets of variates},
  author    = {H. Hotelling},
  booktitle = {Breakthroughs in Statistics: Methodology and Distribution},
  pages     = {162--190},
  year      = {1992},
  publisher = {Springer}
}

@article{morcos2018insights,
  title   = {Insights on representational similarity in neural networks with canonical correlation},
  author  = {A. Morcos and others},
  journal = {Advances in Neural Information Processing Systems},
  volume  = {31},
  year    = {2018}
}

@article{ward2011my,
  title   = {My science tutor: A conversational multimedia virtual tutor for elementary school science},
  author  = {W. Ward and others},
  journal = {ACM Transactions on Speech and Language Processing (TSLP)},
  volume  = {7},
  number  = {4},
  pages   = {1--29},
  year    = {2011}
}

@inproceedings{attia2023kid,
  title={Kid-whisper: Towards bridging the performance gap in automatic speech recognition for children vs. adults},
  author={A. Attia and others},
  booktitle={Proc. AAAI/ACM Conference on AI, Ethics, and Society},
  year={2024}
}

@inproceedings{ogi_kids,
  author       = {Khaldoun Shobaki and
                  John{-}Paul Hosom and
                  Ronald A. Cole},
  title        = {The {OGI} kids{\({^2}\)} speech corpus and recognizers},
  booktitle    = {{INTERSPEECH}},
  year         = {2000}
}

@misc{kendall2023coraal,
  author       = {Kendall, Tyler and Farrington, Charlie},
  title        = {The {C}orpus of {R}egional {A}frican {A}merican {L}anguage},
  year         = {2023},
  version      = {2023.06},
  doi          = {10.7264/1ad5-6t35},
  url          = {https://doi.org/10.7264/1ad5-6t35}
}

@inproceedings{cmu_benchmark,
 author = {Ying, Anyu and others},
 booktitle = {Workshop on Child Computer Interaction - WOCCI},
 title = {{Benchmarking Training Paradigms, Dataset Composition, and Model Scaling for Child ASR in ESPnet}},
 year = {2025}
}

@inproceedings{fan2024benchmarking,
  author={R. Fan and others},
  title={Benchmarking Children's ASR with Supervised and Self-supervised Speech Foundation Models},
  booktitle={{INTERSPEECH}},
  year={2024}
}

@inproceedings{ChenH0A0RLTA25,
  author    = {Li{-}Wei Chen and others},
  title        = {Exploring Prediction Targets in Masked Pre-Training for Speech Foundation Models},
  booktitle    = {{ICASSP}},
  year         = {2025}
}

@misc{NIST-SCTK,
  author = {Fiscus, J.G.},
  title = {{SCTK: The NIST Scoring Toolkit}},
  year = {2007},
  organization = {National Institute of Standards and Technology},
  note = {[Software]},
  accessdate = {February 15, 2025}
}

@inproceedings{RekeshKKMNHHPKBG23,
  author    = {Dima Rekesh and others},
  title     = {Fast Conformer With Linearly Scalable Attention For Efficient Speech Recognition},
  booktitle = {{ASRU}},
  year      = {2023}
}

@misc{nvidia_canary_qwen_25b,
  title        = {{Canary-Qwen-2.5B}},
  author       = {{NVIDIA}},
  howpublished = {\url{https://huggingface.co/nvidia/canary-qwen-2.5b}},
  year         = {2025},
  note         = {Hugging Face model},
}

\end{document}